\documentclass[prb,superscriptaddress,twocolumn,showpacs]{revtex4}
\usepackage{graphicx}
\usepackage{dcolumn}
\usepackage{bm}

\usepackage{array}
\newcolumntype{L}[1]{>{\raggedright\let\newline\\\arraybackslash}p{#1}}
\newcolumntype{C}[1]{>{\centering\let\newline\\\arraybackslash}p{#1}}

\newcommand{\lbco}{$\rm La_{2- \it x}Ba_{\it x}CuO_{4}$}

\newcommand{\ybco}{$\rm YBa_2Cu_3O_{y}$}
\newcommand{\ybcosff}{$\rm YBa_2Cu_3O_{6.44}$}

\newcommand{\lsim}{{ _< \atop ^\sim}}

\newcommand{\pla}{$\rm CuO_2$}

\usepackage{color}  
\definecolor{gray}{rgb}{0.4,0.4,0.4}
\definecolor{gray}{rgb}{0,0,0}

\newcommand{\cut}[1] {\textcolor{green}{[----]}}

\begin{document}

\title{Competing charge, spin, and superconducting orders in underdoped $\bf YBa_{2}Cu_{3}O_{y}$}

\author{M. H\"ucker}
\affiliation{Condensed Matter Physics \&\ Materials Science Department, Brookhaven National Laboratory, Upton, New York 11973, USA}

\author{N. B. Christensen}
\affiliation{Department of Physics, Technical University of Denmark, DK-2800 Kongens Lyngby, Denmark}

\author{A. T. Holmes}
\affiliation{School of Physics and Astronomy, University of Birmingham, Birmingham B15 2TT, United Kingdom}

\author{E. Blackburn}
\affiliation{School of Physics and Astronomy, University of Birmingham, Birmingham B15 2TT, United Kingdom}

\author{E. M. Forgan}
\affiliation{School of Physics and Astronomy, University of Birmingham, Birmingham B15 2TT, United Kingdom}

\author{Ruixing Liang}
\affiliation{Department of Physics $\&$ Astronomy, University of British Columbia, Vancouver, Canada}
\affiliation{Canadian Institute for Advanced Research, Toronto, Canada}

\author{D. A. Bonn}
\affiliation{Department of Physics $\&$ Astronomy, University of British Columbia, Vancouver, Canada}
\affiliation{Canadian Institute for Advanced Research, Toronto, Canada}

\author{W. N. Hardy}
\affiliation{Department of Physics $\&$ Astronomy, University of British Columbia, Vancouver, Canada}
\affiliation{Canadian Institute for Advanced Research, Toronto, Canada}

\author{O. Gutowski}
\affiliation{Deutsches Elektronen-Synchrotron DESY, 22603 Hamburg, Germany}

\author{M.~v.~Zimmermann}
\affiliation{Deutsches Elektronen-Synchrotron DESY, 22603 Hamburg, Germany}

\author{S. M. Hayden}
\affiliation{H. H. Wills Physics Laboratory, University of Bristol, Bristol, BS8 1TL, United Kingdom}

\author{J. Chang}
\affiliation{Institut de la Materiere Complexe, Ecole Polytechnique de Lausanna (EPFL), CH-1015 Lausanne, Switzerland}

\date{\today}

\begin{abstract}
To explore the doping dependence of the recently discovered charge density wave (CDW)
order in YBa$_{2}$Cu$_{3}$O$_{y}$, we present a bulk-sensitive high-energy x-ray study
for several oxygen concentrations, including strongly underdoped \ybcosff . Combined with
previous data around the so-called 1/8 doping, we show that bulk CDW order exists at
least for hole concentrations ($p$) in the \pla\ planes of $0.078 \lsim p \lsim 0.132$.
This implies that CDW order exists in close vicinity to the quantum critical point for
spin density wave (SDW) order. In contrast to the pseudogap temperature $T^*$, the onset
temperature of CDW order decreases with underdoping to $T_{\rm CDW}\sim 90$~K in \ybcosff
. Together with a weakened order parameter this suggests a competition between CDW and
SDW orders. In addition, the CDW order in \ybcosff\ shows the same type of competition
with superconductivity as a function of temperature and magnetic field as samples closer
to $p=1/8$. At low $p$ the CDW incommensurability continues the previously reported
linear increasing trend with underdoping. In the entire doping range the in-plane
correlation length of the CDW order in $b$ axis direction depends only very weakly on the
hole concentration, and appears independent of the type and correlation length of the
oxygen-chain order. The onset temperature of the CDW order is remarkably close to a
temperature $T^\dagger$ that marks the maximum of $1/(T_1T)$ in planar $^{63}$Cu NQR/NMR
experiments, potentially indicating a response of the spin dynamics to the formation of
the CDW. Our discussion of these findings includes a detailed comparison to the charge
stripe order in \lbco .
\end{abstract}

\pacs{74.72.-h, 61.05.cp, 71.45.Lr, 74.25.Dw}

\maketitle

\section{Introduction}

The recent discovery of charge ordered ground states in Y, Bi, La and Hg-based high
temperature superconductors emphasizes the need to understand the competition between
these states and superconductivity in underdoped
cuprates.~\cite{Wu11a,Ghiringhelli12a,Chang12a,Wu13b,SilvaNeto14a,Comin14a,Christensen14a,
Thampy14a,Croft14a,Tabis14a} One of the outstanding questions is how these states are
related to the Fermi surface topology. Quantum oscillation experiments on the archetypal
bi-layer system \ybco\ (YBCO) indicate a reconstruction of the large Fermi surface
typical for overdoped cuprates, into one with small Fermi pockets near a hole
concentrations of $p\sim 1/8$.~\cite{DoironLeyraud07a,Bangura08a,Singleton10a,
Sebastian12a,Vignolle11a,Millis07a,Harrison12a,Vojta12a} Similar quantum oscillation
measurements in a single-layer Hg-based cuprate provide further evidence that Fermi
pockets are a common property around this so-called
1/8-anomaly.~\cite{Barisic13a,DoironLeyraud13a}
A change from positive to negative Hall and Seebeck coefficients in YBCO around this
doping region led to the interpretation that the Fermi pocket must have electron like
character.~\cite{LeBoeuf07a,LeBoeuf11a,Laliberte11a,Chang10a,Hackl10a} Negative Seebeck
and Hall coefficients are also observed in La$_{1.8-x}$Eu$_{0.2}$Sr$_x $CuO$_4$ (Eu-LSCO)
and several other La-based cuprates~\cite{Sera89,Nakamura92,Huecker98b,Noda99a,Suzuki02b,
Li07c,Chang10a,Laliberte11a} that are known to exhibit charge and spin-stripe
order~\cite{Christensen14a,Thampy14a,Croft14a,Kivelson03a,Tranquada95a,Fujita04a,
Huecker07b,Fink09a,Wilkins11a,Huecker11a,Huecker12a,Katano00a,Lake02a,Wu12a}. This
strongly suggested that charge and/or spin order may exist in YBCO as
well.~\cite{LeBoeuf07a,Taillefer09a,Chang10a,Laliberte11a} Evidence of charge order was
indeed revealed by NMR~\cite{Wu11a}, x-ray
diffraction~\cite{Chang12a,Ghiringhelli12a,Achkar12a,Blackburn13a,BlancoCanosa13a}, and
ultrasound~\cite{LeBoeuf13a} experiments. However, the identified wave vectors have been
linked to a two-$\bf q$ charge density wave (CDW) order from Fermi surface nesting rather
than stripe order. In both Bi- and Hg-based cuprates the ordering wave vector was found
to approximately match a nesting vector that connects the tips of the Fermi arcs,
providing further support for a nesting
scenario.~\cite{SilvaNeto14a,Comin14a,Tabis14a,Harrison14a}

\begin{table*}
\caption{\label{tab1} Characteristic properties of the studied \ybco\ single crystals:
oxygen content $y$, structure of oxygen-chain order, superconducting transition
temperature $T_c$, hole content $p$, sample size, onset temperature $T_{\rm CDW}$ and
incommensurability $\delta_b$ at $T_c$ of the CDW order, resolution corrected correlation
lengths $\xi_b$ of the CDW order and the chain order at $T_c$ in the direction of the $b$
axis. The $\xi_b$(CDW) value for $y=6.54$ and the $T_{\rm CDW}$ values for $y=6.54$,
$6.67$, and $6.75$ were taken from Refs.~\onlinecite{Chang12a,Blackburn13a}. $\xi_b$(CDW)
was measured at ${\bf Q} = (0,\delta_b,6.5)$. $\xi_b$(chain) was measured at ${\bf Q} =
(0.5, 0, 6)$, $(0.375, 0, 6)$, and $(0.333, 0, 6.5)$ for o-II, o-VIII, and o-III,
respectively.} \small
\begin{ruledtabular}
\begin{tabular}{lcccccC{0.4cm}C{0.1cm}C{0.4cm}C{0.1cm}C{0cm}cccccccc}
$y$ in & oxygen & $T_c$ & hole        & & & \multicolumn{5}{c}{sample size  }                  & & & $T_{\rm CDW}$ &  $\delta_b$ & $\xi_b$(CDW) & & & $\xi_b$(chain)  \\
YBCO   & order  & (K)   & content $p$ & & & \multicolumn{5}{c}{$a \times b \times c$ (mm$^3$)} & & &   (K)         & (r.l.u.)    & ($\rm \AA$)  & & & ($\rm \AA$)     \\ \hline
6.44   & o-II   & 42    & 0.078       & & & 1.45 & $\times$ & 1.68 & $\times$ & 0.46           & & &   90(15)      &  0.337(2)   &    51(7)     & & &  169(10)        \\
6.512  & o-II   & 59    & 0.096       & & & 2.2  & $\times$ & 1.46 & $\times$ & 0.25           & & &   145(10)     &  0.331(2)   &    61(7)     & & &  233(10)        \\
6.54   & o-II   & 58    & 0.104       & & & 3.1  & $\times$ & 1.9  & $\times$ & 0.16           & & &   155(10)     &  0.328(2)   &    66(7)     & & &  --             \\
6.67   & o-VIII & 67    & 0.123       & & & 3.1  & $\times$ & 1.7  & $\times$ & 0.6            & & &   140(10)     &  0.315(2)   &    63(7)     & & &  138(10)        \\
6.75   & o-III  & 74    & 0.132       & & & 3.5  & $\times$ & 1.8  & $\times$ & 0.5            & & &   140(10)     &  0.305(4)   &    64(10)    & & &  116(10)        \\
6.92   & o-I    & 93    & 0.165       & & & 1.91 & $\times$ & 1.81 & $\times$ & 0.57           & & &    --         &   --        &    --        & & &  --             \\
\end{tabular}
\end{ruledtabular}
\end{table*}

In spite of this tremendous progress, the connection between some of the observations
remains unclear. Doping experiments may thus provide a powerful tool for further tests.
Several recent studies on YBCO indicate a significant qualitative change of the
electronic properties at a critical doping of approximately $p_c\sim
0.08$.~\cite{Dai01a,LeBoeuf11a,Laliberte11a,Haug09a,Haug10a,Sebastian12a,Baek12a} In
particular, the absence of a negative Seebeck and Hall effect below $p_c$ suggests a
disappearance of the proposed electron pocket, which has motivated explanations in terms
of a Lifshitz transition, i.e, a transition that involves a change of the Fermi surface
topology.~\cite{LeBoeuf11a,Laliberte11a,Norman10a,Norman10b} The region below $p_c$ also
exhibits a low temperature one-$\bf{q}$ spin-density-wave (SDW)
order~\cite{Hinkov08a,Stock05a}, and an electronic liquid crystal state at higher
temperatures~\cite{Haug10a}, which are reminiscent of the spin stripe phase in La-based
cuprates.~\cite{Fujita02a} There is no obvious relationship between the SDW order below
$p_c$ and the CDW order above $p_c$. In fact in YBCO, magnetic excitations are gapped in
the doping region where quantum oscillations and CDW have been observed so
far.~\cite{Hinkov07a,Li08a} This shows that a detailed knowledge of the doping dependence
of the CDW order is critical. NMR and x-ray studies have identified charge order down to
approximately $p = 0.104$.~\cite{Ghiringhelli12a,Blackburn13a,Wu13b} Hence, it is still
an open question how the CDW order evolves as the critical point $p_c \sim 0.08$ is
approached.

Here we report a high energy x-ray diffraction study of the CDW order in two underdoped
ortho-II YBCO crystals with $y=6.44$ ($T_c=42$~K) and $y=6.512$ ($T_c=59$~K), as well as
an optimally doped crystal with $y=6.92$ ($T_c=93$~K); see Tab.~\ref{tab1}. While both
underdoped crystals exhibit CDW order, no evidence of this order was found for $y=6.92$.
Much of the attention will concentrate on the results for $y=6.44$, with those for
$y=6.512$ being very similar to our previous data for $y=6.54$.~\cite{Blackburn13a} The
hole concentration~\cite{Liang06a} of the \ybcosff\ crystal is $p \sim 0.078$ and hence
it is in close vicinity to the above mentioned quantum critical point for SDW
order~\cite{Haug10a}, and the proposed Lifshitz transition~\cite{LeBoeuf11a}. The CDW
order in that crystal is weakened but significant and can be traced up to $T_{\rm
CDW}\sim 90$~K. Upon cooling below $T_c$ the CDW reflection is partially suppressed, but
can be enhanced by a magnetic field applied perpendicular to the \pla\ planes. The
ordering wave vector of the CDW in \ybcosff\ continues the growing trend versus
underdoping previously identified around 1/8-doping.~\cite{Blackburn13a}

Two further findings may shed light on the nature of the CDW order in YBCO. First, its
correlation length $\xi_b$ along the $b$ axis, i.e., parallel to the chains, shows no
dependence on the oxygen order in the chain layers, and varies only weakly as a function
of doping, which may indicate that local properties play a role. Second, we find a
remarkable agreement between the CDW onset temperature $T_{\rm CDW}$ and a temperature
$T^\dagger$ below which $1/(T_1T)$ decreases in planar $^{63}$Cu NQR/NMR
experiments.~\cite{Takigawa91a, Berthier97a,Auler97a,Baek12a,Wu13a} We argue that the
opening of a CDW gap may influence the planar Cu spin dynamics. The derived phase
diagrams strongly indicate that CDW order not only competes with SC, but also with SDW
order. Finally, we discuss differences and similarities of the CDW order in YBCO and the
charge stripe order in \lbco\ (LBCO).

\section{Experimental details}
\begin{figure*}
\center{\includegraphics[width=0.8\textwidth]{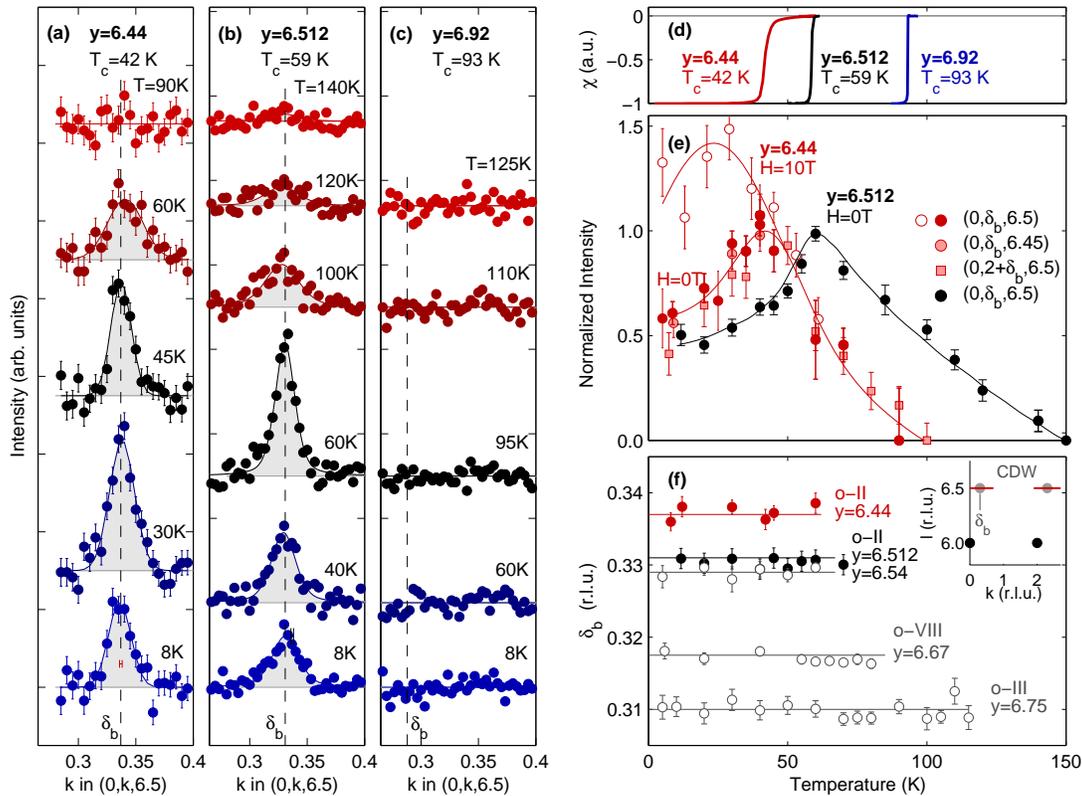}} \caption{(color online)
Temperature dependence of the CDW order in YBCO. (a,b) $k$ scans at zero magnetic field
through ${\bf Q} = (0,\delta_b,6.5)$ for ortho-II crystals with $y=6.44$ and $6.512$,
showing that the CDW order vanishes into the background noise at $T\sim 90$~K and
$\sim$145~K, respectively. Solid lines are least-squares fits using a Gaussian line
shape. Vertical dashed lines indicate the incommensurability $\delta_b$. (c) $k$ scans
for the optimally doped crystal with $y=6.92$ reveal no evidence of a CDW peak. The
dashed line in (c) indicates an estimated CDW peak position based on a linear
extrapolation of the doping dependence $\delta_b(y)$ in Fig.~\ref{fig4}(d).
Sloping backgrounds have been subtracted from all scans that are shifted vertically for
clarity. The red horizontal bar in (a) at $T=8$~K indicates a typical transverse
resolution full width at half maximum. (d) Normalized diamagnetic susceptibility of the
three crystals, showing sharp SC transitions; see Table~\ref{tab1}. (e) Normalized
intensity of the CDW reflections at ${\bf Q} = (0,\delta_b,6.5)$ and $(0,2+\delta_b,6.5)$
versus temperature at zero magnetic field ($H=0$~T) for $y=6.44$ and $6.512$, as well as
at $H=10$~T for $y=6.44$. (f) $\delta_b$ versus temperature for five different dopings.
The data sets are limited to temperatures where $\delta_b$ could be reliably determined.
The inset shows a section of the reciprocal space $(0,k,\ell)$ with the trajectories of
typical $k$ scans through the CDW peaks at ${\bf Q} = (0,\delta_b,6.5)$ and
$(0,2+\delta_b,6.5)$. Solid lines in (e,f) are guides to the eye.}\label{fig1}
\end{figure*}

Synthesis, oxygen annealing, and detwinning procedures for the YBCO single crystals were
described in Ref.~\onlinecite{Liang12a}, and resulted in sharp SC transitions; see
Fig.~\ref{fig1}(d). This indicates well-defined carrier concentrations of $p\sim 0.078$
($y=6.44$), $0.096$ ($y=6.512$), and $0.165$ ($y=6.92$).~\cite{Liang06a} The two
underdoped ortho-II samples are $\sim$99$\%$ detwinned. In addition to these three new
crystals we have re-measured two crystals, used in our previous studies, with $p\sim
0.123$ ($y=6.67$), and $0.132$ ($y=6.75$).~\cite{Blackburn13a,Chang12a}
The high energy x-ray diffraction experiments were carried out with triple-axis
instruments at beamline P07 at PETRA III, DESY, and beamline 6-ID-D at the Advanced
Photon Source (APS) at Argonne National Laboratory. The beam size varied between
0.5$\times$0.5 and 1$\times$1~$\rm mm^2$, and the photon energy was set to E$_{\rm ph}$ =
80~keV. The rectangular crystals, with dimensions listed in Tab.~\ref{tab1}, were mounted
with the $(0,k,\ell)$ zone in the scattering plane, and studied in bulk sensitive
transmission geometry. Two different sample environments were used: a closed cycle
cryostat reaching $T\sim 7$~K, and a magnet cryostat allowing temperatures down to 3~K
and magnetic fields up to $H=10$~T along the $c$ axis of the crystals. Scattering vectors
${\bf Q} = (h,k,\ell)$ are specified in units of $(2\pi/a,2\pi/b,2\pi/c)$ of the
orthorhombic unit cell with space group Pmmm.
The correlation lengths of the CDW order and the oxygen-chain order in the direction of
the $b$ axis are defined by $\xi_b = (\textrm{HWHM} \times b^*)^{-1}$, where HWHM is the
half-width at half-maximum of the corresponding superstructure reflection.
The results are compared to our previously published work for $y=6.54$, $6.67$, and
$6.75$, obtained under similar or identical conditions at beamlines BW5 at DORIS III,
DESY, and P07.

\section{Results}
\subsection{Temperature dependence}
\label{temp}
The CDW order in YBCO leads to weak satellite reflections at wave vectors ${\bf Q} =
\boldsymbol{ \tau} + {\bf q}_{\rm CDW}$ where ${\bf q}_{\rm CDW} = (\delta_a, 0, 0.5)$
and $(0, \delta_b, 0.5)$ are the ordering wave vectors, and $\boldsymbol{\tau}$ a
fundamental Bragg reflection.\cite{Chang12a} In Fig.~\ref{fig1}(a,b) we show $k$ scans
through the position ${\bf Q} = (0,\delta_b,6.5)$ for the two ortho-II compositions
$y=6.44$ and $6.512$ at different temperatures. Both crystals clearly display a CDW
reflection, which makes $y=6.44$ the composition with the currently lowest reported hole
concentration with CDW order. In contrast, no evidence of a CDW peak is observed for
$y=6.92$ in Fig.~\ref{fig1}(c) in the area of the estimated peak position.

\begin{figure*}
\center{\includegraphics[width=0.85\textwidth]{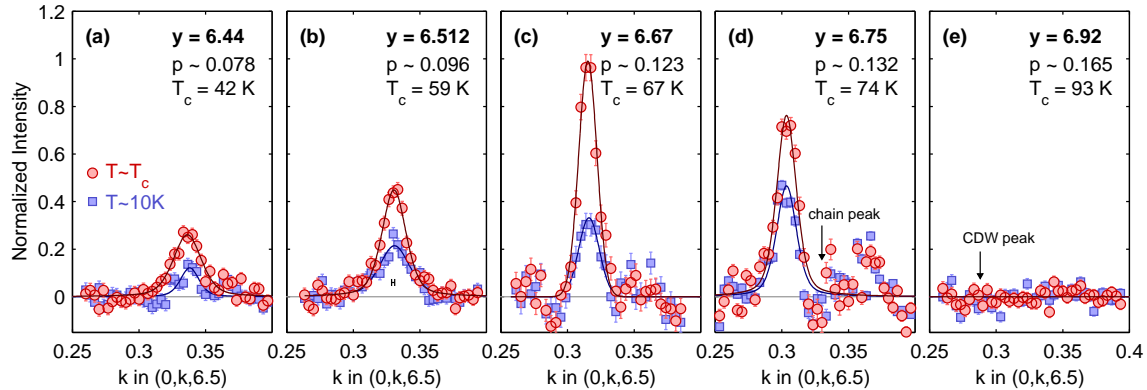}} \caption{(color online) Doping
dependence in zero magnetic field of the CDW peak intensity in YBCO. (a-e) $k$ scans at
$T \sim T_c(y)$ (circles) and $T \sim 10$~K (squares) through the CDW reflection ${\bf Q}
= (0,\delta_b,6.5)$ for the oxygen concentrations $y=6.44$, 6.512, 6.67, 6.75, and 6.92.
All intensities are displayed after background subtraction and careful normalization to
reflect changes as a function of doping (see text for details). Solid lines through the
peaks are least-squares fits using a Gaussian line shape. The horizontal bar in (b)
indicates a typical resolution (full width at half maximum). The arrow in (d) marks the
position of a subtracted peak from the ortho-III oxygen order in the minority twin
domain, which is responsible for the lower statistics in that area. The arrow in (e)
shows an estimated CDW peak position as explained in Fig.~\ref{fig1}. Examples of data
including background counts are given in
Refs.~\onlinecite{Chang12a,Blackburn13a}.}\label{fig2}
\end{figure*}

Recent x-ray diffraction studies (resonant~\cite{BlancoCanosa13a} and
non-resonant~\cite{Blackburn13a}) on ortho-II ordered YBCO crystals with $y\sim6.54$
demonstrated a two-$\bf q$ structure of the CDW order. However, in both cases strongly
anisotropic structure factors are observed, with CDW satellites along ${\bf a}^*$ being
generally sparser and weaker than along ${\bf b}^*$. In addition the background from the
tails of oxygen ordering peaks is larger along ${\bf a}^*$. A similar situation is found
for the ortho-II crystal with $y=6.44$. For the ortho-II crystal with $y=6.512$ no
measurements of equivalent CDW peaks in the $(h,0,\ell)$ zone have been conducted yet. In
this paper, we therefore focus on CDW reflections found along ${\bf b}^*$.

As can be seen in Fig.~\ref{fig1}(e) the temperature dependence for zero magnetic field
of the CDW peak intensity for $y=6.44$ and $6.512$ is similar to that previously reported
for higher dopings.~\cite{Blackburn13a,Ghiringhelli12a,Chang12a,BlancoCanosa13a} In the
normal state the intensity grows smoothly upon cooling, reaches a maximum at $T_c$, and
then is substantially suppressed in the SC state. For $y=6.44$, this dependence was
consistently measured at ${\bf Q}=(0,\delta_b,6.5)$ and $(0,2+\delta_b,6.5)$.
A major difference concerns the onset temperature $T_{\rm CDW} \sim 90$~K of the CDW
order for $y=6.44$, which is about 50~K lower than for $y \geq
6.512$.~\cite{Blackburn13a} Finally, in Fig.~\ref{fig1}(f) we show that the
incommensurability $\delta_b$ for $y=6.44$ and $6.512$ fits well into the existing doping
dependence and is approximately independent of temperature for all $y$.

\subsection{Doping dependence}
\label{doping}
Next we turn to the doping dependence of the CDW order for zero magnetic field in
Fig.~\ref{fig2}. For all samples, scans were performed on the CDW reflection ${\bf Q} =
(0,\delta_b,6.5)$. Because $6.5 c^* \gg \delta_b b^*$, scans along $k$ benefit from the
excellent transverse resolution indicated in Fig.~\ref{fig2}(b). After lining up on the
nearest Bragg reflection $(0,0,6)$, it is thus straightforward to measure the
incommensurability $\delta_b$ and the correlation lengths $\xi_b$ of the CDW order with
high accuracy; see inset of Fig.~\ref{fig1}(f) and Fig.~\ref{fig4}(d,e).
On the other hand, it is much harder to extract the doping dependence of intensities. For
this purpose, we have remeasured five samples -- all mounted on the same sample holder --
in a single experiment. The data were normalized in two different ways which led to very
similar results: (i) a direct normalization of all intensities by the incident x-ray
flux, probed sample volume, and absorption effects, and (ii) a normalization by the
integrated intensity of the $(0,0,2)$ Bragg reflection~\cite{bragg}, which accounts for
the same factors as (i) and is shown in Fig.~\ref{fig2}.

For conventional CDW systems, the resulting integrated intensities, $I$, are proportional
to the square of the CDW order parameter $\Delta$, i.e., $\sqrt{I} \propto \Delta {\rm
(CDW)}$.~\cite{Jaramillo09a} We would like to normalize $\Delta {\rm (CDW)}$ so that its
maximum value in the YBCO system is 1.
Due to the competition with superconductivity, the zero field, zero temperature value of
$\Delta {\rm (CDW)}$ is less than 1 for all $y$.
However, in the limit $T\rightarrow0$ and in a magnetic field $H$ approaching the upper
critical field $H_{c2}$, it is conceivable to assume $\Delta {\rm (CDW)} \sim 1$. For
ortho-VIII YBCO with $p=0.123$, CDW intensities have been measured up to 17
Tesla.~\cite{Chang12a} This field scale is comparable to $H_{c2}\sim 25$~T reported for
this doping.~\cite{Ando02b,Chang11a,Chang12b,Ramshaw12a,Grissonnanche14a} The quantity
$\sqrt{I(p,T,H)}/\sqrt{I(p=0.123,2~\textrm{K},17~\textrm{T})}$ is therefore a good
approximation of the doping, temperature, and magnetic field dependence of $\Delta$(CDW).

The extracted $\Delta{\rm (CDW)}$ values at zero magnetic field and $T \sim T_c(p)$ as
well as $T \sim 10$~K are plotted in Fig.~\ref{fig4}(c) versus hole content $p$. One can
see that $\Delta{\rm (CDW)}$ exhibits a broad maximum at 1/8-doping, and at $T_c$ reaches
about 75\% of its high field value at 2~K.~\cite{Chang12a} As a function of underdoping
$\Delta{\rm (CDW)}$ drops further to about 50\% at $T_c$ and 28\% at 10~K at the critical
point $p_c \sim 0.08$.

Although this clear weakening of the CDW order, as the SDW phase is approached, suggests
a competition between the two phases, the data do not support a complete disappearance of
CDW order at $p_c$. Instead, it suggests a region below $p_c$ where CDW and SDW orders
may overlap. To demonstrate this, Fig.~\ref{fig4}(c) also shows the volume fraction of
the SDW order measured by $\mu$SR.~\cite{Sanna04a} At the hole content of our \ybcosff\
crystal the $\mu$SR data suggest a magnetically ordered volume fraction of 25\%. The true
extent of the overlap depends of course sensitively on the accuracy to which the doping
concentration $p$ is determined for the $\mu$SR and x-ray experiments. Furthermore, it is
well known that the lack of perfect oxygen order at such low oxygen concentrations
results in weak sample inhomogeneity.~\cite{Liang12a,Zimmermann03a}

\begin{figure}[b]
\center{\includegraphics[width=0.49\textwidth]{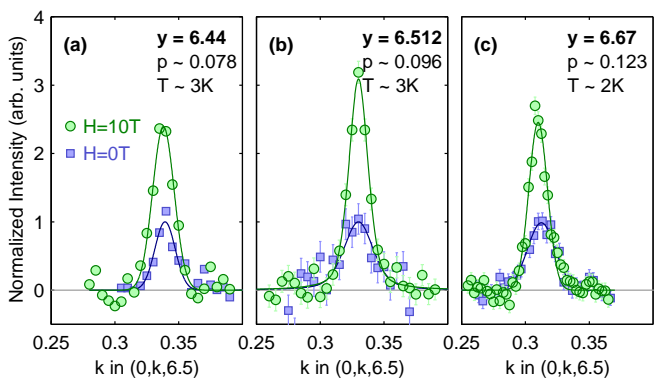}} \caption[]{(color online)
Magnetic field effect on the CDW peak intensity in YBCO at base temperature for
concentrations $y=6.44$, 6.512, and 6.67. (a-c) $k$ scans at $H=0$ and $10$~T through
${\bf Q} = (0,\delta_b,6.5)$. For each doping scans have been normalized by the maximum
peak intensity in zero magnetic field. Solid lines are least-squares fits using a
Gaussian line shape. Sloping backgrounds have been subtracted from all scans. In (a)
error bars are within symbol size.}\label{fig3}
\end{figure}

A question is therefore whether the weak CDW order in \ybcosff\ originates from regions
with $p>0.078$ due to inhomogeneity of the hole concentration. There are several facts
that speak against this scenario. As can be seen in Fig.~\ref{fig1}(e) the CDW intensity
peaks right at $T_c = 42$~K. This implies that CDW and SC compete in those parts of the
sample where $T_c = 42$~K and, therefore, $p \sim 0.078$. We arrive at the same
conclusion based on the doping dependence of the incommensurability $\delta_b$ in
Fig.~\ref{fig1}(f) and Fig.~\ref{fig4}(d). The fact that $\delta_b$ in \ybcosff\
continues the approximately linear doping trend already reported in
Ref.~\onlinecite{Blackburn13a} proves that, in those regions with CDW order, $p$ must be
close to the estimated value. Finally, Fig.~\ref{fig4}(e) shows the correlation length
$\xi_b$ measured at $T \sim T_c$. Obviously, $\xi_b$ varies only weakly with doping.
Although $\xi_b$ for $y=0.44$ is slightly lower than for dopings closer to $p=1/8$, we
would expect it to be significantly shorter, if the CDW order were a minority phase. All
factors taken together, we conclude that CDW order in \ybcosff\ is not a result of sample
inhomogeneity. Thus, we have demonstrated that intrinsic CDW order exists all the way
down to the lower quantum critical point at $p_c$, where it touches and very likely
overlaps with the competing SDW phase.~\cite{Haug10a}

\subsection{Magnetic field dependence}
\label{field}
When suppressing SC with a magnetic field of $H=10$~T applied along the $c$ axis, a
significant enhancement of the CDW peak is achieved, as is shown in Fig.~\ref{fig1}(e)
for $y=6.44$. The slight drop in intensity below 25~K reflects the fact that 10~T is
below the critical field $H_{c2}$ for $y=6.44$ and, thus, insufficient to fully suppress
SC.~\cite{Grissonnanche14a} This high field $T$-dependence is very similar to previous
observations near 1/8-doping.~\cite{Blackburn13a,BlancoCanosa13a,Chang12a}
To compare the doping dependence of the field effect, we show in Fig.~\ref{fig3} data for
the oxygen concentrations $y=6.44$, 6.512, and 6.67 at $T=3$~K and $H=0$ and 10~T.
All scans were performed at ${\bf Q} = (0,\delta_b,6.5)$, and have been normalized by the
peak intensities in zero field. Independent of the hole content, the application of
$10$~T along the $c$ axis enhances the CDW peak by a factor of 2.5 to 3. On an absolute
scale as in Fig.~\ref{fig2} this means that gains are most significant for $p \sim 1/8$.
This seems to correlate with the fact that $H_{c2}$ is minimum at $p =
1/8$.~\cite{Grissonnanche14a}

\section{Discussion}
\subsection{Competing CDW and SDW orders near $p_c$}
The underdoped part of the YBCO phase diagram is complex and interesting because several
electronic phases co-exist with superconductivity; see Fig~\ref{fig5}. The one-$\bf q$
SDW order identified by neutron scattering for dopings just above the critical
concentration $p \sim 0.05$ of the antiferromagnetic phase, vanishes again in vicinity of
the quantum critical point $p_c \sim 0.08$.~\cite{Haug10a} We note that $p_c$ is well
inside the SC dome as well as the ortho-II phase, which both set in at $p \sim 0.05$
($y\sim 6.3$).~\cite{Ando04a,Sebastian12a,Zimmermann03a} For $p>p_c$ superconductivity
was shown to compete with CDW
order.~\cite{Wu11a,Ghiringhelli12a,Chang12a,Achkar12a,LeBoeuf13a,Wu13b} In approximately
the same doping region centered at $p\sim 1/8$, quantum oscillation
experiments~\cite{DoironLeyraud07a, Vignolle11a,Sebastian12a} in concert with high-field
Hall and thermopower measurements~\cite{Laliberte11a,Chang10a} were interpreted in terms
of an electron pocket. So far, CDW order is the most natural explanation for a Fermi
surface reconstruction that produces these pockets.~\cite{Wu11a,Vojta12a}

\begin{figure*}[t]
\center{\includegraphics[width=0.92\textwidth]{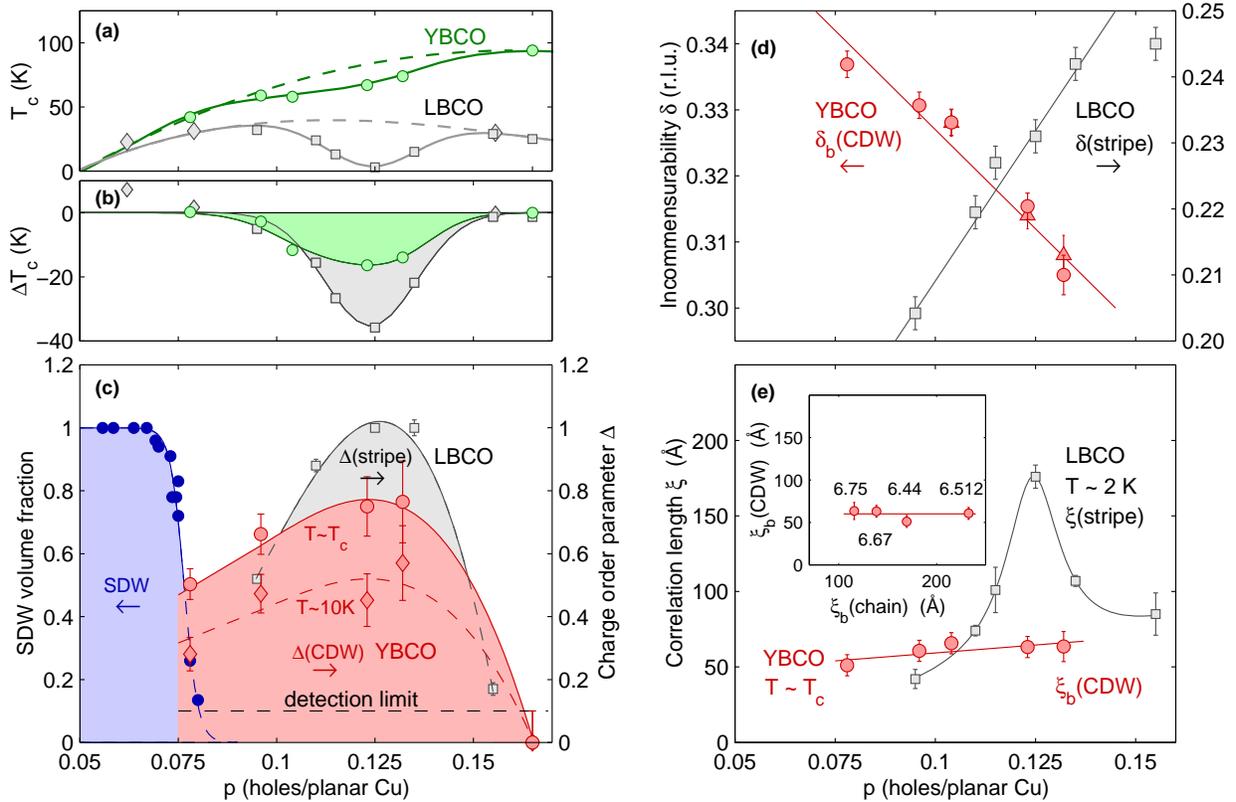}} \caption[]{(color online)
Comparison of CDW order in YBCO with charge stripe order in LBCO as a function of planar
hole concentration $p$ for zero magnetic field. (a) Superconducting transition
temperature $T_c$ and (b) suppression of $T_c$ through 1/8-effect for YBCO from this work
(circles) and Ref.~\onlinecite{Liang06a} (green lines), and for LBCO from
Ref.~\onlinecite{Yamada92} (diamonds) and Ref.~\onlinecite{Huecker11a} (squares). The
dashed gray line in (a) for LBCO is a cubic fit~\cite{cubicfit} of $T_c(p)$ outside the
1/8-region to describe the envelope of the SC dome, while the solid line includes a
Gaussian term to account for the 1/8-anomaly. (c) The right ordinate shows the CDW order
parameter $\Delta{\rm (CDW)}$ in YBCO measured at ${\bf Q} = (0,\delta_b,6.5)$ in zero
magnetic field, and normalized to the high-field low-temperature value of the ortho-VIII
crystal with $y=6.67$ (see text for details). Red circles were measured at $T\sim
T_c(p)$, and red diamonds at $T\sim 10$~K. The solid and dashed red lines are guides to
the eye. The left ordinate represents the SDW volume fraction measured by $\mu$SR (closed
blue circles).\cite{Sanna04a} Gray squares indicate the charge stripe order parameter
$\Delta{\rm (stripe)}$ in LBCO measured with x-rays in zero magnetic field at $T\sim
3$~K.~\cite{Huecker13a} The horizonal dashed line indicates an approximate detection
limit for the high energy x-ray diffraction experiment. (d) CDW incommensurability
$\delta_b$(CDW) in YBCO measured at $T\sim T_c(p)$ and ${\bf Q} = (0,\delta_b,6.5)$ (red
circles) as well as data from Ref.~\onlinecite{Blackburn13a} (red triangles). Gray
squares indicate the charge stripe incommensurability $\delta$(stripe) in
LBCO.~\cite{Huecker11a,Huecker13a} (e) CDW correlation length $\xi_b$(CDW) in YBCO
measured at $T\sim T_c(p)$ and ${\bf Q} = (0,\delta_b,6.5)$ (red circles), and stripe
correlation length $\xi$(stripe) in LBCO~\cite{Huecker11a,Huecker13a} at $T\sim 3$~K
(gray squares). The resolution has been deconvolved, although it is basically negligible;
see Fig.~\ref{fig2}(b) and Tab.~\ref{tab1}. The inset shows $\xi_b$(CDW) of the CDW order
versus $\xi_b$(chain) of the oxygen order, both measured in direction of the $b$ axis.
(d,e) The data for $\delta_b$(CDW) and $\xi_b$(CDW) are average values obtained from
measurements of the same peak ${\bf Q} = (0,\delta_b,6.5)$ in several beam times; see
Tab.~\ref{tab1}.}\label{fig4}
\end{figure*}

\begin{figure}[t]
\center{\includegraphics[width=0.48\textwidth]{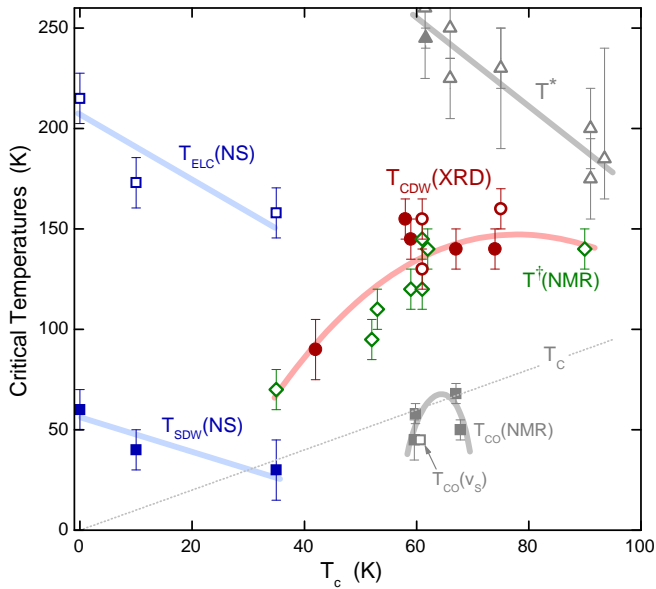}} \caption[]{(color online)
Critical temperatures of competing spin and charge orders in YBCO as a function of $T_c$.
By plotting all data versus $T_c$, ambiguities of plots versus the planar hole content
$p$, due to different ways and difficulties of determining $p$, can be
reduced.~\cite{Liang06a} We show the transition temperatures $T_{\rm CDW}$ of CDW order
measured by high energy x-rays~\cite{Chang12a,Blackburn13a} (closed circles) and soft
x-ray scattering~\cite{Ghiringhelli12a,BlancoCanosa13a,Bakr13a} (open circles), $T_{\rm
CO}$ of charge order detected by high field NMR~\cite{Wu11a,Wu13a} and high field sound
velocity ($v_S$) measurements~\cite{LeBoeuf13a}, $T^\dagger$ determined from the maximum
in $1/(T_1T)$ of planar $^{63}$Cu NQR/NMR as explained in the
text~\cite{Takigawa91a,Berthier97a,Auler97a,Baek12a,Wu13a} (open diamond), the pseudo gap
temperature $T^*$ as detected by means of Nernst effect (open triangle) and resonant
ultrasound measurements~\cite{Daou10a,Shekhter13a} (closed triangle), $T_{\rm SDW}$ of
SDW order (closed blue square) and $T_{\rm ELC}$ of a so called electronic liquid crystal
state as determined by neutron scattering~\cite{Hinkov07a,Haug10a} (open blue square).
$T_c$ is indicated by a dashed line. All solid lines are guides to the eye.
}\label{fig5}
\end{figure}

To make further progress it is obviously critical to understand the region around $p_c$
where CDW crosses over to SDW. If CDW order is connected to the presence of electron
pockets, one would naively expect it to weaken significantly across $p_c$. Our results
would support such a scenario. First, the data for $y=6.44$ and $6.512$ confirm that CDW
order evolves systematically with underdoping, and persists all the way to $p_c \sim
0.08$. Second, the CDW order for $y=6.44$ is weakened, although not as drastically as we
had expected, and the onset temperature $T_{\rm CDW}$ is substantially reduced. Derived
phase diagrams of both the order parameters in Fig.~\ref{fig4}(c), and the onset
temperatures in Fig.~\ref{fig5} strongly indicate a competition between SDW and CDW
phases, which may include a not insignificant region of coexistence. This suggests that
the proposed Lifshitz transition at $p_c$ may occur when the CDW order weakens through
phase competition.

\subsection{CDW onset temperature}
The decrease of $T_{\rm CDW}$ with underdoping is an important observation, because other
characteristic temperatures in the underdoped regime, especially the pseudogap
temperature $T^*$, appear to continue to increase with
underdoping.~\cite{Fauque06a,Mook08a,Xia08a,Daou10a,LeBoeuf11a,Shekhter13a,Tabis14a} To
identify properties potentially connected to the CDW order, Fig.~\ref{fig5} shows a
critical temperature $T^\dagger$ that marks a broad maximum in the $1/(T_1T)$ signal of
planar $^{63}$Cu NQR/NMR
experiments.~\cite{Takigawa91a,Berthier97a,Auler97a,Baek12a,Wu13a} The agreement between
$T_{\rm CDW}$ and $T^\dagger$ is very suggestive. This NQR/NMR feature at $T^\dagger$ is
characteristic for samples in the pseudogap phase where $T_c < T^\dagger < T^*$. It is
apparent that $T^\dagger$ decreases with underdoping, too. The origin of $T^\dagger$ is a
matter of debate, but common interpretations involve the onset of spin freezing, and a
gapping of the low energy spin fluctuations by the pseudogap or by incoherent pairing in
the normal state.~\cite{Wu13a,Hinkov07a,Baek12a,Tdagger} With respect to the incoherent
pairing scenario, it is worth noticing yet another property that shares a similar doping
dependence as $T_{\rm CDW}$ and $T^\dagger$, and that is the onset temperature reported
in Ref.~\onlinecite{Kokanovic12a} of so called precursor diamagnetism~\cite{Li10a} in the
static magnetic susceptibility $\chi(T)$ of YBCO for $H\parallel c$. The discovery of CDW
order surrounding the 1/8-anomaly introduces important aspects to this debate. In
particular, the peak in the relaxation the NMR may indicate a response of the spin
dynamics to the formation of the CDW. Associated effects on static magnetic
susceptibility and electronic transport coefficients are likely. More work is certainly
needed to elucidate such connections. It should be noted that various comparisons of
$T_{\rm CDW}$ to other critical temperatures have been
reported.~\cite{LeBoeuf11a,Chang12a,Shekhter13a,Bakr13a,Tabis14a}

\subsection{CDW order in YBCO vs stripe order in LBCO}
\subsubsection{Order parameter and incommensurability}
The striking similarity of the thermopower response found in YBCO and stripe ordered
La-based cuprates suggests that a reconstruction of the Fermi surface into one with small
electron pockets may be a universal feature of charge ordered
cuprates.~\cite{Laliberte11a,Li07c,Noda99a} It is therefore interesting to compare the
doping evolution of the charge orders in YBCO and the prototypical stripe compound
LBCO.~\cite{Thampy13a,Huecker11a,Huecker13a} To this end, Fig.~\ref{fig4}(a) displays
$T_c(p)$ of both systems, clearly showing the well-known suppression of $T_c$ near
1/8-doping.~\cite{Liang06a,Yamada92,Huecker11a} To quantify the 1/8-anomaly in YBCO, the
authors of Ref.~\onlinecite{Liang06a} have subtracted $T_c(p)$ from a fit of the envelope
of the superconducting dome. Here we do the same for LBCO and plot the difference $\Delta
T_c(p)$ for both systems in Fig.~\ref{fig4}(b).~\cite{cubicfit} One can see that LBCO
compared to YBCO shows a stronger suppression of $T_c$. This agrees well with the fact
that LBCO also shows the larger charge order parameter; see Fig.~\ref{fig4}(c). At
1/8-doping charge stripe order in LBCO is already fully developed in zero magnetic
field~\cite{Huecker11a,Huecker13a}, while in YBCO the zero-field CDW order is incomplete
and, thus, SC not fully suppressed~\cite{Liang06a,LeBoeuf11a}.

The different doping dependence in YBCO and LBCO of the charge order incommensurability
has already been pointed out~\cite{Blackburn13a} but is repeated in Fig.~\ref{fig4}(d) to
put the new values for $y=6.44$ and $6.512$ into perspective. One can see that
$\delta_b$(CDW) continues the approximately linear doping trend around 1/8-doping all the
way down to $p=0.078$. In the stripe phase the incommensurabilities of the charge and
spin orders are coupled, whereas those of the CDW and SDW orders in YBCO seem to be
unrelated.~\cite{BlancoCanosa13a, Blackburn13a,Huecker11a} If one considers the doping
dependence of $\delta_b$(SDW) of the SDW order in YBCO, it appears that this order might
actually be a relative of the stripe order in La-based cuprates.~\cite{Fujita02a,Haug10a}
In this respect it is interesting that Zn doping in YBCO causes a weakening of the CDW
state (and as a matter of fact a suppression of the broad maximum of $1/(T_1T)$ in the
planar $^{63}$Cu NQR/NMR~\cite{Julien00,Timusk99a}) as well as the reappearance of a SDW
state at dopings $p\sim 1/8$.~\cite{BlancoCanosa13a,Suchaneck10a} This shows that the CDW
and SDW orders not only compete with SC, but also with each other. The results for the
hole doping dependence of the SDW and CDW phases near $p_c$ in Fig.~\ref{fig4}(c) and
Fig.~\ref{fig5} support the same idea.

\subsubsection{Correlation length}
Another interesting difference between YBCO and LBCO concerns the doping dependence of
the in-plane charge order correlation length $\xi$. As can be seen in Fig.~\ref{fig4}(e)
the correlation length of the charge stripe order in \lbco\ exhibits a pronounced maximum
at $p=1/8$ of $\xi{\rm (stripe)}\sim 180~{\rm \AA}$, but drops rapidly by a factor of
three within a 3\% variation of $p$. In contrast, in YBCO the correlation length
$\xi_b$(CDW) at $T\sim T_c(p)$ is always quite short and varies only weakly for $0.078
\leq p \leq 0.132$. Moreover, $\xi_b$(CDW) appears to be independent of the type of
oxygen order (ortho-II, VIII, or III), and also independent of the correlation length
$\xi_b$(chain) of the oxygen chain order measured in the same direction $\bf{b}^*$; see
inset in Fig.~\ref{fig4}(e). One could argue that the type of oxygen order should have
little effect on $\xi_b$(CDW), because it only affects the way the chains are arranged
along $\bf{a}^*$. However, a recent study also finds no effect of the oxygen order on the
CDW correlation length along $\bf{a}^*$.~\cite{Achkar13b} A weak maximum of $\xi{\rm
(CDW)}$ at $p\sim 0.12$ has been indicated in Ref.~\onlinecite{Tabis14a}, which is less
apparent from our data set where each point was obtained in an identical way. At an
average we find $\xi_b {\rm (CDW)} \sim 60~{\rm \AA}$ at $T_c$, which is comparable to
$\xi{\rm (stripe)}$ in LBCO far away from 1/8-doping. Similar correlation lengths have
been found in several soft x-ray studies on YBCO, with a weak tendency toward slightly
larger values, which may be due to the smaller probed sample
volume.~\cite{Ghiringhelli12a,Achkar13a,BlancoCanosa13a,Thampy13a}

The situation is comparable at base temperature in the SC state and zero magnetic field,
because $\xi_b{\rm (CDW)}$ does not change significantly below $T_c$, as can be seen in
Fig.~\ref{fig1}(a,b) and several other
studies.~\cite{Chang12a,Ghiringhelli12a,BlancoCanosa13a} Only when suppressing
superconductivity with a magnetic field, can $\xi_b {\rm (CDW)}$ be increased below
$T_c$. Nevertheless, even for $p\sim 1/8$ and almost complete suppression of
superconductivity, $\xi_b {\rm (CDW)}$ does not exceed $\sim 100$~${\rm
\AA}$.~\cite{Chang12a} On the one hand this shows that the coexistence with
superconductivity is one of the factors that limit $\xi_b {\rm (CDW)}$ in YBCO. On the
other hand, the independence of $\xi_b{\rm (CDW)}$ from the chain superstructures may
indicate that local physics plays an important role as well, as will be discussed below.

\subsection{CDW order and oxygen chain order}
The above differences between the charge order superstructures in LBCO and YBCO are not
unexpected because of the materials' distinct crystal structures. The absence of chain
layers in La-214 materials is certainly the most important difference. In YBCO these
chains introduce an orthorhombic distortion that breaks the four-fold rotational symmetry
of the \pla\ planes, which in itself could stabilize a charge
order.~\cite{Kivelson98,Huecker10a,Vojta12a,Achkar12a} This is quite similar to LBCO
where charge stripe order is most stable in the low-temperature tetragonal (LTT) phase
which breaks the rotational symmetry of the individual planes as
well.~\cite{Fujita04a,Kim08a,Huecker11a} Interestingly, the correlation length
$\xi$(stripe) of the charge stripe order in LBCO appears unrelated to $\xi$(LTT) of the
LTT phase.~\cite{Huecker10a} Here we have shown that the same is true for $\xi_b$(CDW)
and $\xi_b$(chain) in YBCO; cf. Fig.~\ref{fig4}(e). In both systems charge order does not
seem to couple in a simple way to the long range structure that breaks the rotational
symmetry. In fact, in LBCO with $p=0.125$ charge stripes even form when the long range
ordered LTT phase is absent, i.e., by restoring a four-fold rotational symmetry of the
planes at high pressures.~\cite{Huecker10a} In this high-symmetry phase it was found that
$\xi$(stripe) actually matches $\xi$(LTT) of persisting diffuse peaks from a quenched
disorder of local LTT-type distortions.~\cite{Huecker10a,Fabbris13a}

Therefore, one might speculate whether in YBCO the CDW order in the planes couples to
local rather than long range properties of the chains. In general a coupling of the
electronic correlations in the planes and the chains has been a matter of intense
debate.~\cite{Derro02a,Yamani06a,Baek12a,Achkar13a,Achkar13b,Wu14a} One of the reasons is
that the chains are prone to 1D like Peierls instabilities.~\cite{Gruener94a} In fact
several scanning tunneling microscopy (STM) studies have identified a modulation of the
local density of states along the chains, i.e., along the $b$
axis.~\cite{Derro02a,Maki02a} In agreement with that, a recent soft x-ray angle-resolved
photoemission spectroscopy (SX-ARPES) experiment detected a gapped surface chain band
whose nesting vector matches the modulation wave vector found by
STM.~\cite{Zabolotnyy12a} Comparing our bulk sensitive x-ray data results to these
surface related observations is not straightforward, since the chain layer at the surface
is known to be heavily overdoped.~\cite{Hossain08a} However, both the modulation period
($\sim$9-14~\AA) and the correlation length ($\sim 40 $~\AA) reported by STM and SX-ARPES
studies~\cite{Derro02a,Maki02a,Zabolotnyy12a} are intriguingly close to our values for
$\delta_b{\rm (CDW)}$ and $\xi_b$(CDW) in Fig.~\ref{fig4}. Common interpretations of the
charge modulations on the chains are Friedel oscillations~\cite{Derro02a,Maki02a} caused
by chain defects, and a Peierls-like CDW instability~\cite{Zabolotnyy12a}. The
correlation length of the Friedel oscillations, being a local perturbation, may not
depend strongly on hole doping or $\xi_b$(chain). Thus, the almost independence of $\xi_b
{\rm (CDW)}$ observed in our x-ray study is at least not inconsistent with a coupling of
the planar CDW order to quenched disorder states on the chains. A recent NMR study
arrives at similar conclusions.~\cite{Wu14a} This discussion shows that both scenarios, a
coupling of the planar CDW order to the symmetry breaking potential of well ordered
chains, as well as to local chain properties deserve further consideration.

\section{Conclusions}
In summary, we have identified CDW order in underdoped YBCO with $y=6.44$ and $6.512$
using high energy x-ray diffraction. Strong emphasis was placed on the first sample with
a hole content $p=0.078$ that is very close to the critical point $p_c$. The CDW of this
crystal shows the same competition with superconductivity as a function of temperature
and magnetic field as previously reported around $p=1/8$
doping.\cite{Chang12a,Ghiringhelli12a,Achkar12a,Blackburn13a,BlancoCanosa13a} However,
onset temperature and order parameter of the CDW order are significantly reduced. This
implies that CDW also competes with the SDW phase, which becomes the dominant state
competing with superconductivity below $p_c \sim 0.08$.~\cite{Haug10a} A detailed
comparison of the doping dependence of the CDW order in YBCO and the charge stripe order
in LBCO is presented. One striking difference is that the correlation length of the CDW
order is relatively short ($\sim 60$~\AA) and almost independent of $p$, whereas in the
case of charge stripe order it shows a pronounced maximum reaching $\sim 180$~\AA\ at
$p=1/8$.~\cite{Huecker11a} Among potential scenarios we consider a coupling between the
CDW order in the planes and local states in the chains.~\cite{Derro02a,Zabolotnyy12a}
Furthermore, we find an interesting agreement between the CDW onset temperature and a
temperature in nuclear resonance experiments that marks a maximum in the planar
relaxation rate.~\cite{Takigawa91a,Berthier97a,Auler97a,Baek12a,Wu13a} We argue that the
maximum may indicate a response of the low energy spin fluctuations to the formation of
the CDW. When plotted versus a common $T_c$-scale, our results for \ybcosff\ with
$T_c=42$~K are still slightly above the highest-$T_c$ sample (35~K) with confirmed SDW
order,~\cite{Haug10a} and slightly below the lowest-$T_c$ samples with gapped magnetic
excitations (48~K)~\cite{Li08a} and quantum oscillations (54~K)~\cite{Sebastian10a}. This
clearly emphasizes the need for additional doping experiments. Overall, our results show
that the CDW phase exists in a broad doping region approximately congruent with that
characterized by negative Hall and Seebeck coefficients, thus providing additional
support for a potential connection between the CDW order and electron like Fermi
pockets.~\cite{LeBoeuf11a,Laliberte11a}

\section{Acknowledgements}
We acknowledge fruitful discussions with J. M. Tranquada, L. Taillefer, H.-J. Grafe, A.
Achkar, and W. Ku. This work was supported by the Office of Basic Energy Sciences (BES),
Division of Materials Science and Engineering, U.S. Department of Energy (DOE), under
Contract No. DE-AC02-98CH10886; the Danish Agency for Science, Technology and Innovation
under DANSCATT; the EPSRC (Grants No. EP/G027161/1, No. EP/J015423/1, and No.
EP/J016977/1); the Wolfson Foundation; the Royal Society; and the Swiss National Science
Foundation through NCCR-MaNEP and Grant No. PZ00P2-142434. Use of the Advanced Photon
Source, an Office of Science User Facility operated for the U.S. DOE Office of Science by
Argonne National Laboratory, was supported by the U.S. DOE under Contract No.
DE-AC02-06CH11357.


\begin{thebibliography}{105}
\bibitem[{bra()}]{bragg}
\bibinfo{note}{To normalize a superstructure reflection it is common to use a
  near by fundamental Bragg reflection to maintain a similar scattering
  geometry. For the CDW peak at ${\bf Q} = (0,\delta_b,6.5)$ that would be
  $(0,0,6)$. However, if the Bragg reflection is very strong, as is the case
  for $(0,0,6)$, extinction effects can introduce a significant error to the
  normalization. Therefore, we have used the much weaker $(0,0,2)$ Bragg
  reflection. At photon energies on the order of $E_{\rm ph} = 80$~keV this is
  possible, because scattering angles are small, and the scattering geometries
  for $(0,0,2)$ and $(0,0,6)$ not very different.}
\bibitem[{cub()}]{cubicfit}
\bibinfo{note}{We note that the envelope of the SC dome of LBCO is asymmetric,
  and cannot be fit with a quadratic function centered at optimum doping
  ($p=0.16$) as is the case in YBCO. Instead we have used a third order
  polynomial that peaks around $p=0.12$.}
\bibitem[{Tda()}]{Tdagger}
\bibinfo{note}{In general planar $^{63}$Cu NQR/NMR experiments are suited to
  sort out whether the decrease of $1/(T_1T)$ below $T^\dagger$ is caused by
  spin freezing or the gapping of low energy spin fluctuations. Spin freezing
  becomes dominant at low doping and causes a broadening and weakening of the
  low temperature spectra, which is the so-called wipe out effect. This is not
  the case at higher hole concentrations and temperatures
  $T_c<T<T^\dagger$.~\cite{Baek12a} However, in the intermediate doping range
  it can be difficult to distinguish these effects. Relaxation rates may drop
  because of a gap, or because spins, fluctuating at frequencies that cause
  strong relaxations, freeze out first. In Fig.~\ref{fig4} we show a
  significant number of points between $T_c=52$~K and $62$~K, which we count
  among the high-doping regime, and clearly show a decrease of $T^\dagger$ with
  underdoping. According to Ref.~\onlinecite{Wu13a}, in the sample with
  $T_c=35$~K and $T^\dagger=70$~K, which is most affected by spin freezing,
  $T^\dagger$ is assumed to still be above the onset temperature of the wipe
  out effect.}
\end{thebibliography}

\end{document}